# Designing the Interactive Memory Archive (IMA): A Socio-Technical Framework for AI-Mediated Reminiscence and Cultural Memory Preservation

Ron Fulbright

University of South Carolina Upstate
Spartanburg, SC 29303, USA
fulbrigh@uscupstate.edu

**Abstract.** This paper introduces the Interactive Memory Archive (IMA), a conceptual framework for AI-mediated reminiscence designed to support cognitive engagement among older adults experiencing memory loss. IMA integrates multimodal sensing, natural language conversational scaffolding, and cloud-based archiving within the familiar form of a large format historical picture book. The model theorizes reminiscence as a guided, context-aware interaction eliciting autobiographical memories and preserving them as cultural artifacts. The paper positions IMA as a theoretical contribution, articulates testable propositions, and outlines a research agenda for future empirical, technical, and ethical inquiry.

**Keywords:** Human centered computing → Human computer interaction (HCI); • Applied computing → Health informatics; • Information systems → Digital libraries and archives.

## 1    Introduction

Dementia, age-related memory loss, and cognitive decline represent some of the most significant public health challenges of the twenty first century. Globally, over 55 million suffer with over 7 million adults aged 65 and older living with Alzheimer's dementia in the USA alone  [1, 2, 3]. This is expected to rise as the population ages. The global dementia-care products market is projected to be over $25 billion as of 2026 and growing [4]. The dementia treatment market is expected to grow from $18 billion to over $28 billion by 2039 [5].

Although short term memory is often impaired, long term autobiographical memory frequently remains accessible well into the progression of cognitive decline [6]. A growing body of work demonstrates reminiscence, storytelling, and conversational engagement around long term memories can provide meaningful cognitive and emotional benefits for older adults [7, 8]. Reminiscence Therapy (RT) is a non-pharmacological treatment to help individuals with dementia. These activities not only stimulate neural pathways associated with memory retrieval but also strengthen identity, social connection, and psychological well-being [9]. Recent research has explored using technology such as robots, virtual reality, haptic devices, and keepsake boxes to facilitate reminiscence [10, 11, 12, 13]. "Legacy Capture" platforms have been developed using artificial intelligence to interview the user making it easier to capture stories [14, 15, 16].

Elderly individuals often spend hours paging through historical picture books featuring images of their hometown, local neighborhoods, and landmarks. The images trigger personal recollections, yet the memories they evoke typically vanish once the moment passes. Families may record stories informally (clicking record on a smartphone) but such recordings are difficult to organize, index, or integrate into broader historical or cultural archives. Needed is a way to meaningfully integrate the therapeutic value of reminiscence with the long-term preservation of autobiographical and cultural memory.

This paper introduces the Interactive Memory Archive (IMA) as a conceptual and socio technical framework designed to address this gap. Core constructs and mechanisms of IMA are discussed. IMA is situated within existing research on cognitive aging, human–AI interaction, and digital heritage. Testable propositions to guide future empirical work are proposed. In doing so, the paper positions IMA as a research agenda catalyst—a call for interdisciplinary exploration into how AI companions might support cognitive health, enrich personal and collective memory, and reshape the ways societies preserve lived experience.

IMA is presented not as a finalized technological solution, but as a conceptual provocation. As such, this work should be understood as a position paper articulating a vision, identifying a conceptual gap, and proposing a theoretical model inviting further exploration.

## 2       Background and Theoretical Foundations

Understanding the potential of the Interactive Memory Archive (IMA) requires situating it within several intersecting domains of scholarship: cognitive aging and memory systems, human–AI interaction and companion technologies, and the study of cultural memory and digital archiving.

### 2.1    Cognitive Aging and the Structure of Memory

Research confirms memory decline in dementia is not uniform and follows a temporal pattern called Ribot's Law [17, 18]. Short-term and working memory tend to deteriorate early in the progression of dementia, while long-term autobiographical memory often remains relatively intact. Therapeutic approaches emphasizing reminiscence, narrative reconstruction, and conversational engagement have been developed. Studies show prompting older adults to recall personal experiences can stimulate cognitive activity, reinforce identity, and improve emotional well-being [19, 20]. This type of therapy is called Reminiscence Therapy (RT). Digital storytelling can improve the quality of life for people living in residential aged care [28].

## 2.2    Human–AI Interaction and Reminiscence Therapy

The emergence of conversational AI systems opens new possibilities for supporting older adults. AI companions have been explored as tools for social engagement, emotional support, and cognitive stimulation [21, 22, 23, 24]. Notable systems include: LifeBio [14], an app using AI-Powered prompts to record a person's voice and stories and RecallAid [15], a tool using AI to provide specific triggers to stimulate recall. Remme is an AI-powered RT platform specifically designed for dementia and memory care [29]. Remme mimics a video smartphone call and uses a conversational AI "therapist" to guide people through their memories.

Research shows RT is an effective intervention for individuals with dementia improving cognitive function, mood, and reducing depression with digital RT outperforming traditional formats [30, 31, 32, 33, 34, 35, 36, 37, 38]. Many older adults are hesitant to engage with digital devices, perceiving them as unfamiliar or intrusive. This creates a design challenge: how to embed advanced AI capabilities within a form factor that feels natural, comfortable, and non-technical.

## 2.3    Cultural Memory, Storytelling, and Digital Archiving

Digital archiving initiatives have expanded the capacity to preserve such materials, yet they typically rely on deliberate contributions—written accounts, recorded interviews, or curated collections. StoryCorps has helped over 700,000 people have meaningful conversations about their lives [16]. ForgetMeNot is a mobile app acting as a personalized memory book for people with dementia [26]. ForGetMeNot uses photos, music, and voice recordings to stimulate and capture content able to be shared with others granted access. The National Museum of American History's Stories on Main Street, an online initiative, has collected over 6,000 first-person accounts and over 4,000 photos from across rural and small-town communities in the United States [27]. The American Association of Retired Persons (AARP) suggests telling one's life story is good for health [39].

Autobiographical stories provide insight into local histories, community identities, and lived experiences often absent from formal archives. There is no widely adopted system that captures memories spontaneously as they arise in everyday interactions, nor one linking personal narratives to specific visual artifacts such as historical photographs.

## 2.4    Local, Regional, and Neighborhood Picture Books

Many existing RT frameworks utilize personal items, such as photos, belonging to the subject because personal items stimulate recall of personal memories. However, basing items on local and regional history, common to a number of people, results in more generalizable results and offers a way to scale nationally. Arcadia Publishing has carved out a distinctive niche in American book culture by focusing on local and regional history [40]. Arcadia has published over 12,000 books each focusing on one town or neighborhood containing images of local mills, plants, theaters, churches, and other buildings. Therefore, the Arcadia catalog of images covers thousands of regions of interest in the United States.

## 2.5     Conceptual Gap and Opportunity

No current framework realizes a unified socio-technical system capable of eliciting, capturing, and sharing personal memories at scale. IMA is designed to fill this conceptual gap. By combining multimodal sensing, AI-mediated conversational scaffolding, and cloud-based memory archiving IMA offers a new model for supporting cognitive engagement while simultaneously building a distributed cultural memory repository.

## 3     The Interactive Memory Archive (IMA): Conceptual Model

IMA is proposed as a socio-technical framework integrating cognitive stimulation, AI-mediated conversation, and distributed cultural memory preservation into a unified system. Rather than functioning as a traditional digital tool or therapeutic device, IMA is conceptualized as a *hybrid companion*—a system eliciting, capturing, and sharing autobiographical memories through a familiar interaction with historical imagery [43, 44, 45].

At the heart of IMA is the construct of *reminiscence activation*: the process by which historical images trigger autobiographical recall. This construct draws on cognitive theories of cue-dependent memory retrieval, in which visual stimuli serve as anchors for long-term memory [41, 42]. IMA operationalizes this construct by presenting region-specific historical photographs, like those found in the Arcadia local history books, reliably evoking personal stories, emotions, and associations [40].

### 3.1     System Architecture (Conceptual Overview)

Several embodiments of IMA are envisioned including a smartphone app, a tablet/laptop/desktop app, and a Web app (online). In this paper, we describe one embodiment—a large-format historical picture book intentionally designed to be unobtrusive and familiar. The physical form reduces technological intimidation and aligns with the reading habits of many older adults. Embedded within the book are:

- a microcomputer controller
- cameras and microphones
- a speaker system
- wireless connectivity (WIFI)
- local storage for buffering
- sensors for facial and gaze detection

These components operate invisibly, preserving the aesthetic and tactile qualities of a traditional coffee-table book as shown in Fig. 1.

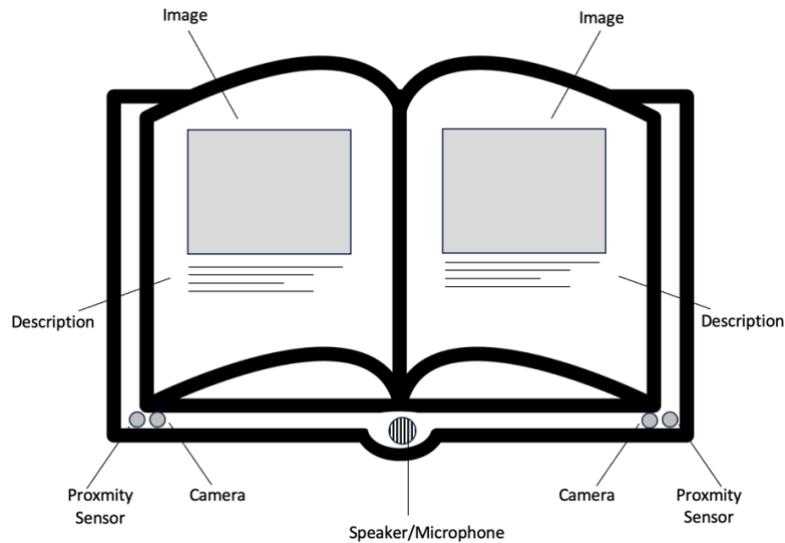

**Fig. 1.** IMA embodied as a large-format historical picture book.

The system architecture supports three core functions: (1) presenting historical images, (2) facilitating conversational engagement, and (3) capturing and transmitting memories to the cloud archive.

### 3.2    Memory Capture and Metadata Linking

A defining feature of IMA is its ability to record narrated memories, link them to structured metadata, and store them in a cloud-based repository accessible by all IMA embodiments. This creates a distributed archive in which personal stories accumulate over time, forming a collective cultural memory spanning individuals, communities, and generations.

While looking at an image, IMA records the narrative, processes the audio, and extracts metadata. Each memory is associated with the specific image being viewed, the identity of the storyteller, and relevant contextual tags. This transforms the constantly evolving memory archive into a durable, searchable artifact able to be accessed by others and preserving the captured personal history for posterity.

The user can record one or more memories for each image. Since many different people will see the image, multiple memories may be associated with each image. The image, being a picture of something with local and regional historic value, is linked to one or more published histories and other descriptive information. These may be textual descriptions IMA can display, or Internet-based resources IMA can link to.

Each interaction with IMA is a session. Of course, an individual is expected to engage in with IMA over an extended period (years hopefully) and therefore experience many sessions with IMA. Figure 2 shows the data schema used by IMA.

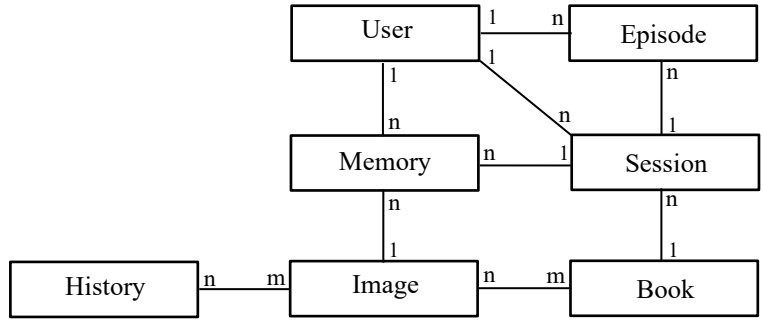

**Fig. 2.** The memory archive data schema.

### 3.3    AI-Mediated Conversational Scaffolding

IMA incorporates an AI-powered natural-language conversational interface (like used by Remme [29]) designed to guide, prompt, and deepen memory recall. This scaffolding is not merely reactive; it is structured to encourage elaboration, clarify details, and sustain engagement. The conversational agent functions as a facilitator.

During conversation about an image, IMA may build rapport by recalling details of previous sessions. Also, elderly with memory loss may not remember they have talked about an image with IMA previously. IMA may choose to use historical information about the contents of the image or memories from previous sessions to enrichen the dialog.

IMA prompts memory recall through a combination of visual cues and conversational nudges. When the user lingers on an image, the system may ask open-ended questions such as "What do you remember about this place?" or "Did you ever visit this neighborhood?" These prompts are grounded in RT principles.

IMA is designed not only to capture memories but also to share them. Since recorded memories from all users are stored in the archive, IMA can make the current user aware of these other memories and even play them upon request. Thus, in subsequent sessions, dialog about a particular image may vary depending on the presence of new, recently recorded, memories and choices made by the user. Reciprocal sharing fosters a sense of connection and community, transforming the system into a living archive of shared experience.

IMA integrates multimodal sensing capabilities such as facial recognition and eye tracking. These mechanisms allow the system to identify the user, determine which image they are viewing, and adapt conversational prompts accordingly. Context awareness ensures the interaction remains grounded in the user's immediate visual and emotional experience.

## 4    Testable Propositions

IMA is a conceptual and theoretical framework. Here, we propose a set of testable propositions guiding future research. These propositions provide a foundation for interdisciplinary inquiry across cognitive science, human–AI interaction, gerontology, and digital heritage studies.

1. **Enhanced Reminiscence Depth**
   Interactions with IMA will elicit more detailed and emotionally rich autobiographical memories than existing RT mechanisms.
2. **Increased Cognitive Activation**
   Regular use of IMA will result in higher levels of cognitive activation.
3. **Improved Emotional Well-Being**
   Users interacting with IMA will report greater emotional satisfaction and reduced feelings of loneliness compared to those using non-interactive materials.
4. **Strengthened Sense of Social Connection**
   Exposure to memories recorded by others—especially from the same region or era—will increase users' perceived sense of community.
5. **Increased Family Engagement**
   Family members will engage more frequently with the memories captured by IMA than with traditional materials.
6. **Unique Cultural Insights**
   Memories captured through IMA will contain historically or culturally significant details not found in formal archives.
7. **Enhanced Archival Value Through Metadata Linking**
   Memories indexed by image, storyteller identity, and thematic content will be more discoverable and more frequently reused by researchers, historians, and community members.
8. **Reduced Cognitive Load Through Familiar Form Factors**
   Users interacting with the book-based embodiment will experience lower cognitive load and higher comfort levels than those interacting with screen-based embodiments.
9. **Emergence of a Distributed Cultural Memory Network**
   Over time, the accumulation of memories across thousands of IMA embodiments will produce a distributed cultural memory network exhibiting identifiable patterns, themes, and regional narratives.

## 5     Conclusion

The Interactive Memory Archive (IMA) is presented as a conceptual and theoretical framework, integrating reminiscence activation, AI-mediated conversational scaffolding, multimodal context awareness, and distributed memory archiving, intended to stimulate new lines of research at the intersection of cognitive aging, human–AI interaction, and cultural memory preservation. IMA opens a broad landscape of empirical, theoretical questions thus is a meaningful contribution to multiple scholarly domains. The constructs and mechanisms described here highlight the potential for AI systems to function not only as tools or companions, but as stewards of lived experience. The testable propositions provide a foundation for future empirical work, offering clear pathways for researchers to evaluate cognitive, emotional, social, and archival impacts. Ultimately, the contribution of this paper lies not in presenting a finished system, but in opening a new line of inquiry. IMA is offered as a vision for how to preserve the stories defining communities and create new forms of cultural memory.